\documentclass[superscriptaddress,aps,prd]{revtex4-1}

\usepackage{graphicx}%
\usepackage{dcolumn}
\usepackage{amsmath,amsthm,amssymb}

\begin{document}

\title{Nonlinear interaction between electromagnetic and gravitational waves: an appraisal}

\author{W. Barreto}
\email{wobarreto@gmail.com}
\affiliation{
Centro de F\'{\i}sica Fundamental, Universidad de Los Andes, M\'erida 5101, Venezuela}

\affiliation{
Departamento de F\'{\i}sica Te\'orica - Instituto de F\'{\i}sica
A. D. Tavares, Universidade do Estado do Rio de Janeiro \\ 
R. S\~ao Francisco Xavier, 524. Rio de Janeiro, RJ, 20550-013, Brazil}

\author{H. P. de Oliveira}\thanks{Corresponding author; hp.deoliveira@pq.cnpq.br}
\affiliation{
Departamento de F\'{\i}sica Te\'orica - Instituto de F\'{\i}sica
A. D. Tavares, Universidade do Estado do Rio de Janeiro \\ 
R. S\~ao Francisco Xavier, 524. Rio de Janeiro, RJ, 20550-013, Brazil}

\author{E. L. Rodrigues}
\email{eduardo.rodrigues@unirio.br}
\affiliation{
Instituto de Bioci\^encias - Departamento de Ci\^encias Naturais, 
Universidade Federal do Estado do Rio de Janeiro\\
Av. Pasteur, 458 - Urca. Rio de Janeiro, RJ, 22290-040, Brazil}
\date{\today}

\begin{abstract}
Wave propagation of field disturbances is ubiquitous. The electromagnetic and gravitational are cousin theories in which the corresponding waves play a relevant role to understand several related physical.  It has been established that small electromagnetic waves can generate gravitational waves and vice versa when scattered by a charged black hole. In the realm of cylindrical spacetimes, we present here a simple nonlinear effect of the conversion of electromagnetic to gravitational waves reflected by the amount of mass extracted from them.

\vspace{10cm}

Awarded Honorable Mention in the Gravity Research Foundation essay competition 2017. 

\end{abstract}

\maketitle

The concept of electromagnetic waves emerged with the modern theory of the electromagnetic field based on the Maxwell equations. Electromagnetic waves propagate with the light velocity and can carry energy, momentum and angular momentum, but the total amount of the charge is conserved. 

On the other hand, gravitational waves are probably the most spectacular prediction of the General Theory of Relativity. The gravitational interaction is a manifestation of spacetime curvature, and the gravitational waves are ripples in the spacetime curvature that propagate with the velocity of light in vacuum. As any other wave, energy, momentum and angular momentum are carried and transferred to particles and electromagnetic fields. Contrary to the electromagnetic waves, gravitational radiation is very tiny, and its direct detection by the consortium LIGO \cite{LIGO} was possible recently after the decades-long effort on the theoretical and technological issues.

The unique aspect of gravitational waves is the mass extraction from the source, contrary to the case of electromagnetic waves in which they do not carry charge from the source. Suppose a bounded distribution of matter that starts to emit gravitational radiation for some time until settling down to a stationary configuration. This configuration is characterized by a smaller mass than the initial since the missing amount of matter is extracted by gravitational waves. Remarkably, Bondi and collaborators \cite{bondi} presented the mathematical tools necessary to understanding more precisely the process of mass extraction by gravitational waves. After a careful analysis of the field equations of an axisymmetric spacetime, they have introduced concepts like the news functions, and the Bondi mass along with their most celebrated result encoded by the Bondi formula.

The theoretical exploration of the relevant aspects of gravitational waves started with the analysis of linearized gravitational waves or waves whose amplitude is tiny allowing to neglect the backreaction of the metric \cite{MTW}. In this case, we can get rid of the nonlinearities and derive a standard wave equation satisfied by the gravitational potentials $h_{\mu\nu}(\mathbf{x},t)$ after a proper gauge choice, where $h_{\mu\nu}(\mathbf{x},t)$ represent the small departure of flat spacetime. In the 50's, Regge and Wheeler \cite{regge_wheeler} studied the stability of the Schwarzschild spacetime. The key element of their research was the behavior of the black hole in interaction with a package or a train of small amplitude gravitational waves.  It turned to be a traditional scattering problem, for which a fraction of the waves is absorbed by the hole whereas and another is radiated away. Later, Gerlach \cite{gerlach} pointed out that in the presence of a background electromagnetic field an electromagnetic wave converts into a gravitational wave and vice versa. Moreover, he advanced that electromagnetic waves scattered by a charged black hole could be converted entirely into gravitational radiation. Olson and Unruh \cite{olson_unruh} looked at the same problem considering odd-parity perturbations in a charged black hole, where they presented analytical estimates and numerical results of the conversion of electromagnetic to gravitational waves.

In this essay, we propose to go beyond the implications of the linear regime of interacting electromagnetic and gravitational waves, by instead, to look into the consequences of the \textit{nonlinear} interaction between both types of waves. To accomplish such task, we have chosen the cylindrical spacetime arena whose general line element is expressed as \cite{kompaneets,jordan}, 

\begin{eqnarray}
ds^2 = -{\rm e}^{2(\gamma - \psi)} (du^2 + 2~du~d\rho) + {\rm e}^{2\psi}(dz+\omega d\phi )^{2} + \rho^{2}{\rm e}^{-2\psi}d\phi ^{2}, 
\end{eqnarray}

\noindent where $u$ is the retarded null coordinate that foliates the spacetime in hypersurfaces $u=\mathrm{constant}$ and $(\rho,z,\phi)$ are the usual cylindrical coordinates. It can be shown that the metric functions $\psi(\rho,u)$ and $\omega(\rho,u)$ represent the two dynamical degrees of freedom of the gravitational field, in which $\psi$ accounts for the polarization mode $+$ while $\omega$ the polarization mode $\times$ \cite{thorne}. The function $\gamma$ plays the role of the gravitational energy of the system; it is connected to the C-energy \cite{stachel,thorne}, more precisely $\gamma (\rho,u)$ gives the total energy per unit length enclosed within a cylinder of radius $\rho$ at the time $u$. 

Stachel \cite{stachel} has extended the analysis of Bondi and collaborators to vacuum cylindrical spacetimes containing both degrees of freedom of the gravitational field. The concepts of news functions, mass aspect are generalized together with the establishment of the Bondi formula. We can step further by including the electromagnetic field as the material content described by the following energy-momentum tensor,

\begin{eqnarray}
T_{\mu\nu} = \frac{1}{4\pi} \left(F_\mu\,^\alpha F_{\nu\alpha} - \frac{1}{4} g_{\mu\nu} F_{\alpha\beta} F^{\alpha\beta}\right)
\end{eqnarray}
 
\noindent where $F_{\mu\nu} = \partial_\mu A_\nu - \partial_\nu A_\mu$ is the electromagnetic field tensor and $A_\mu$ is the potential four-vector. 

The electromagnetic field couples naturally with the gravitational potentials $\psi$ and $\omega$ meaning that, as we have mentioned, electromagnetic waves can generate gravitational waves and the other way round. Instead of presenting the whole set of the field equations we can visualize the coupling between electromagnetic and gravitational fields by the generalized expressions for the Bondi mass and the Bondi formula. Then, from the field equation $R_{\rho\rho}=8\pi T_{\rho\rho}$ together with the definition of the Bondi mass, $M_B(u) = 1/2 \lim_{\rho \rightarrow \infty}\, \gamma(u,\rho)$ \cite{thorne,stachel}, we obtain,

\begin{equation}
M_B(u) = \frac{1}{2}\int_0^\infty\,\left[\rho \left(\frac{\partial \psi}{\partial \rho}\right)^2 + \frac{\mathrm{e}^{4 \psi}}{4 \rho} \left(\frac{\partial \omega}{\partial \rho}\right)^2\right] d \rho + 2\pi \int_0^\infty\,\rho T_{\rho\rho} \,d \rho. 
\end{equation}

\noindent Notice that the last term on the RHS accounts for the contribution of the electromagnetic field to the Bondi mass.

The Bondi mass is not a conserved quantity as the ADM mass, but decay monotonically according to the celebrated Bondi formula which is obtained by taking the asymptotic expression of the combination of the field equations, $R_{uu}-R_{u\rho} = 8 \pi (T_{uu}-T_{u\rho})$,

\begin{equation}
\frac{d M_B}{d u}=-\left[\left(\frac{d c}{d u}\right)_\psi^2+\left(\frac{d c}{d u}\right)_\omega^2 + \left(\frac{d c}{d u}\right)_{EM}^2 \right],\label{eq15}
\end{equation}

\noindent where the first two terms on the RHS are the news functions associated with the degrees of freedom of the gravitational wave, whereas the third term represents the news function related to the electromagnetic potentials. 

Under the action of the news functions, mass-energy of the system is extracted by the combined action of gravitational and electromagnetic waves. We can interpret each term on the RHS of Eq. (4) as \textit{channels from which the mass-energy is carried out}. Let us consider a specific model for which $\omega=0$, i.e. the case of a polarized gravitational wave, and $A_\mu=(0,0,A_2,A_3)$ to be compatible with cylindrical symmetry \cite{thorne,celestino}. The news function associated to the gravitational potential $\omega$ vanishes, while the remaining news functions are, 

\begin{eqnarray}
\left(\frac{d c}{d u}\right)_\psi^2=\lim_{\rho \rightarrow \infty} \rho \left(\frac{\partial \psi}{\partial u}\right)^2,\;\;\, \left(\frac{d c}{d u}\right)_{A_2}^2=\lim_{\rho \rightarrow \infty} \rho \mathrm{e}^{-2\psi}\left(\frac{\partial A_2}{\partial u}\right)^2,\;\;\, \left(\frac{d c}{d u}\right)_{A_3}^2=\lim_{\rho \rightarrow \infty} \frac{\mathrm{e}^{2\psi}}{\rho}\left(\frac{\partial A_3}{\partial u}\right)^2.
\end{eqnarray}


\noindent Moreover, we can infer the total amount of mass extracted during the time interval $\Delta u= u-u_0$ in each channel by evaluating the quantity $I(u)$ defined by,

\begin{eqnarray}
I(u) = \int_{u_0}^u\,\left(\frac{dc}{du}\right)^2 du.
\end{eqnarray}

\noindent We have named $I_\psi(u), I_2(u)$ and $I_3(u)$ the integrals corresponding to each of the news functions.

\begin{figure}[h]
\begin{center}
\includegraphics[scale=0.2]{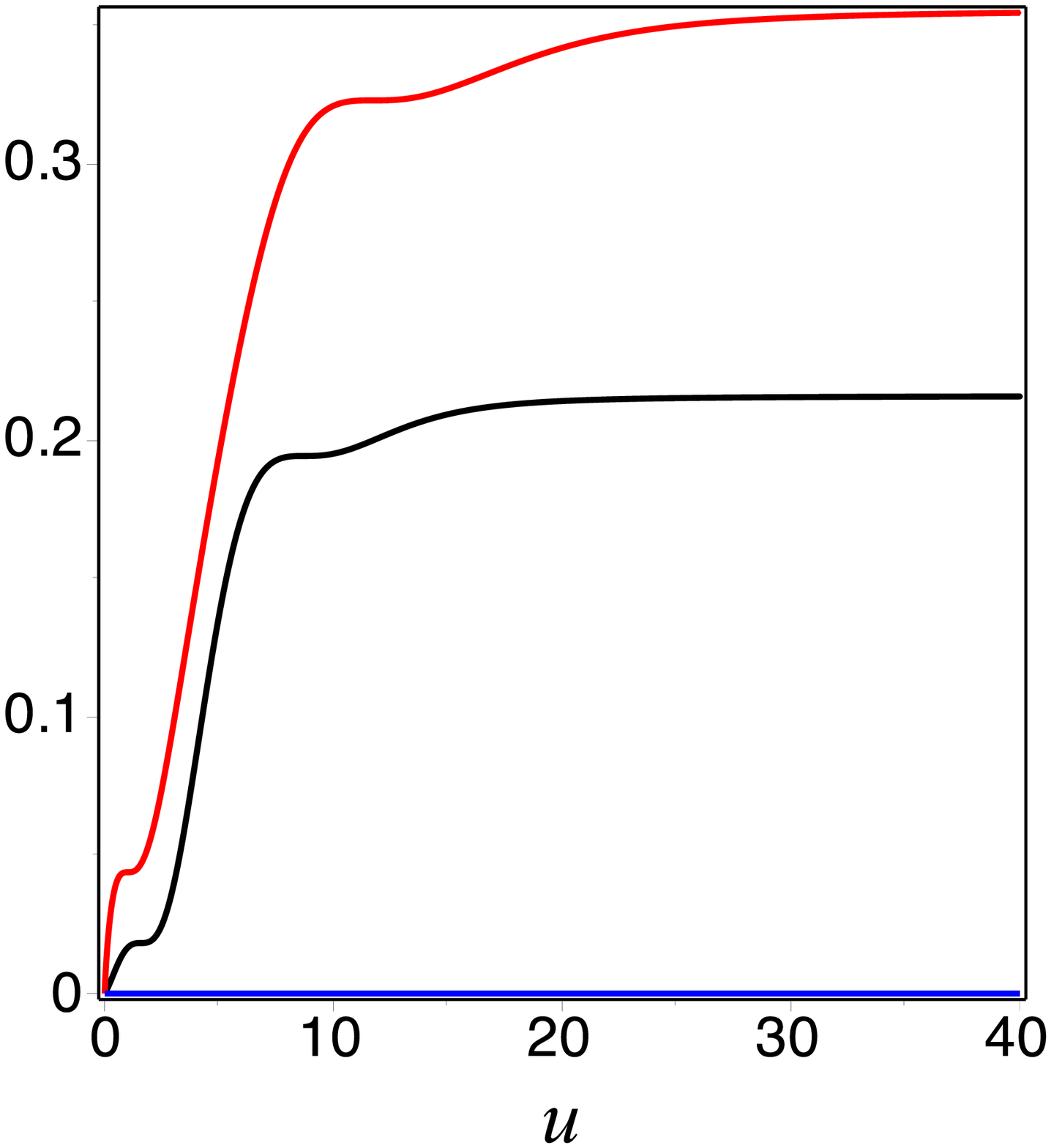}\includegraphics[scale=0.2]{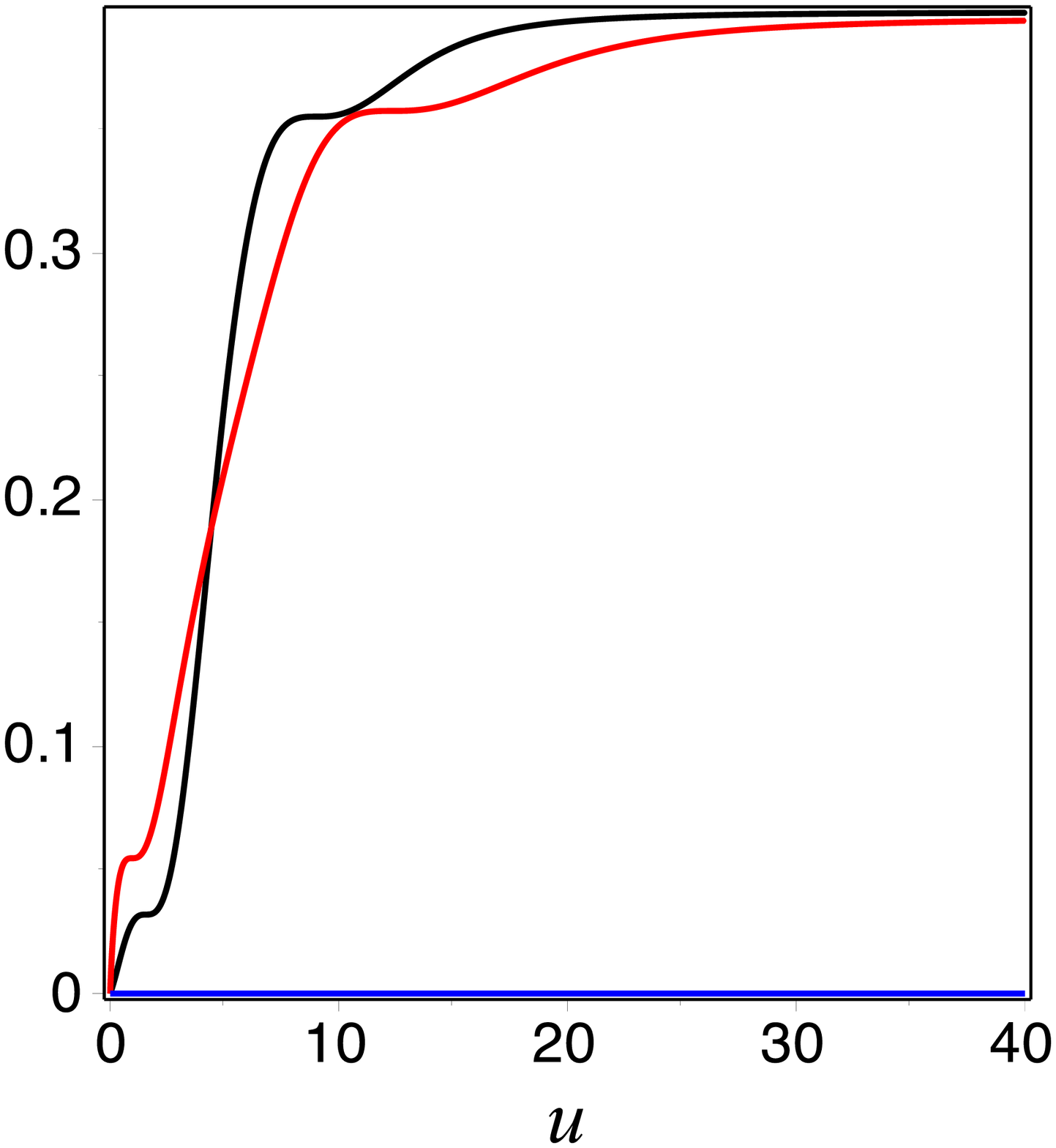}
\includegraphics[scale=0.2]{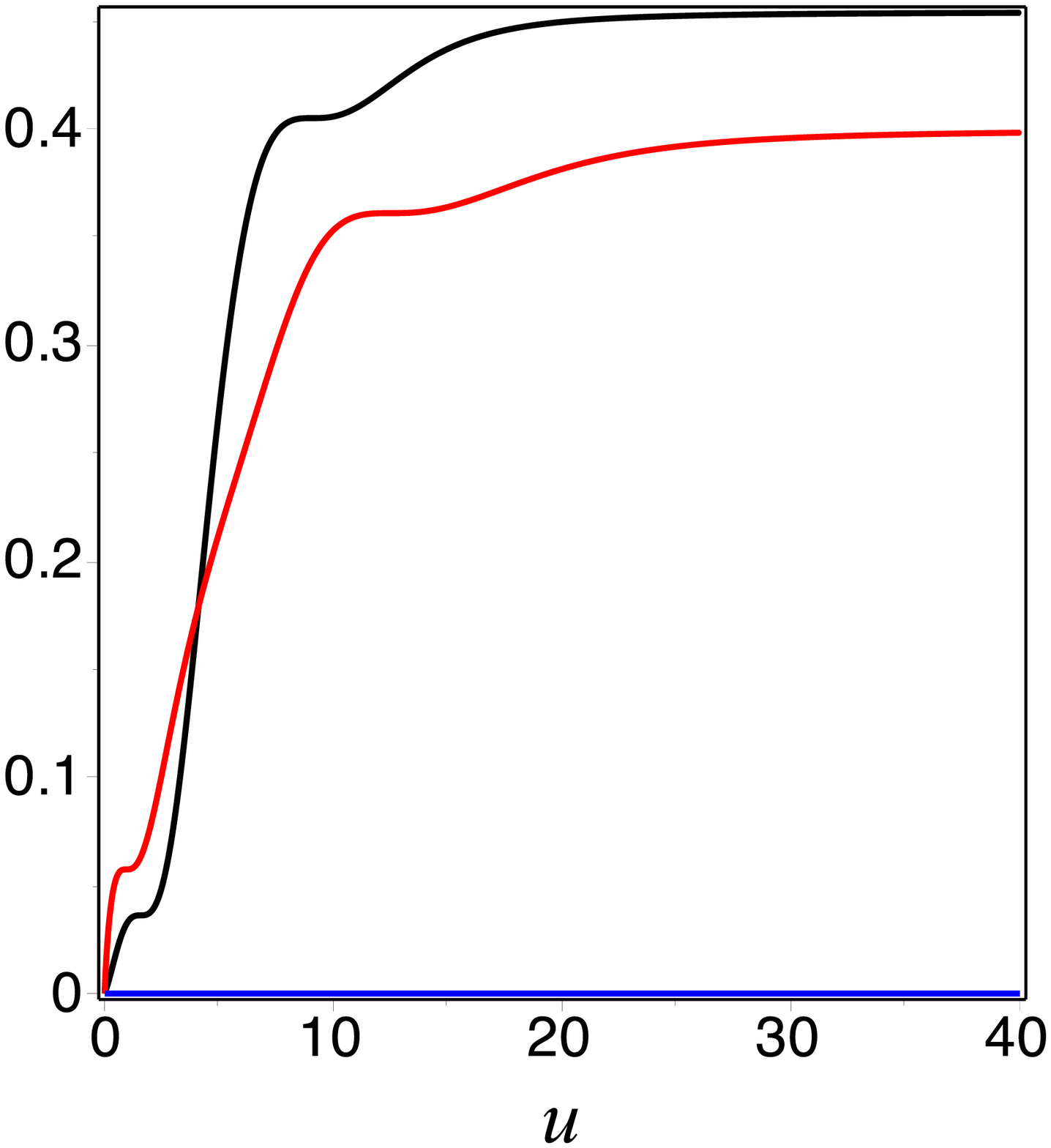}\includegraphics[scale=0.2]{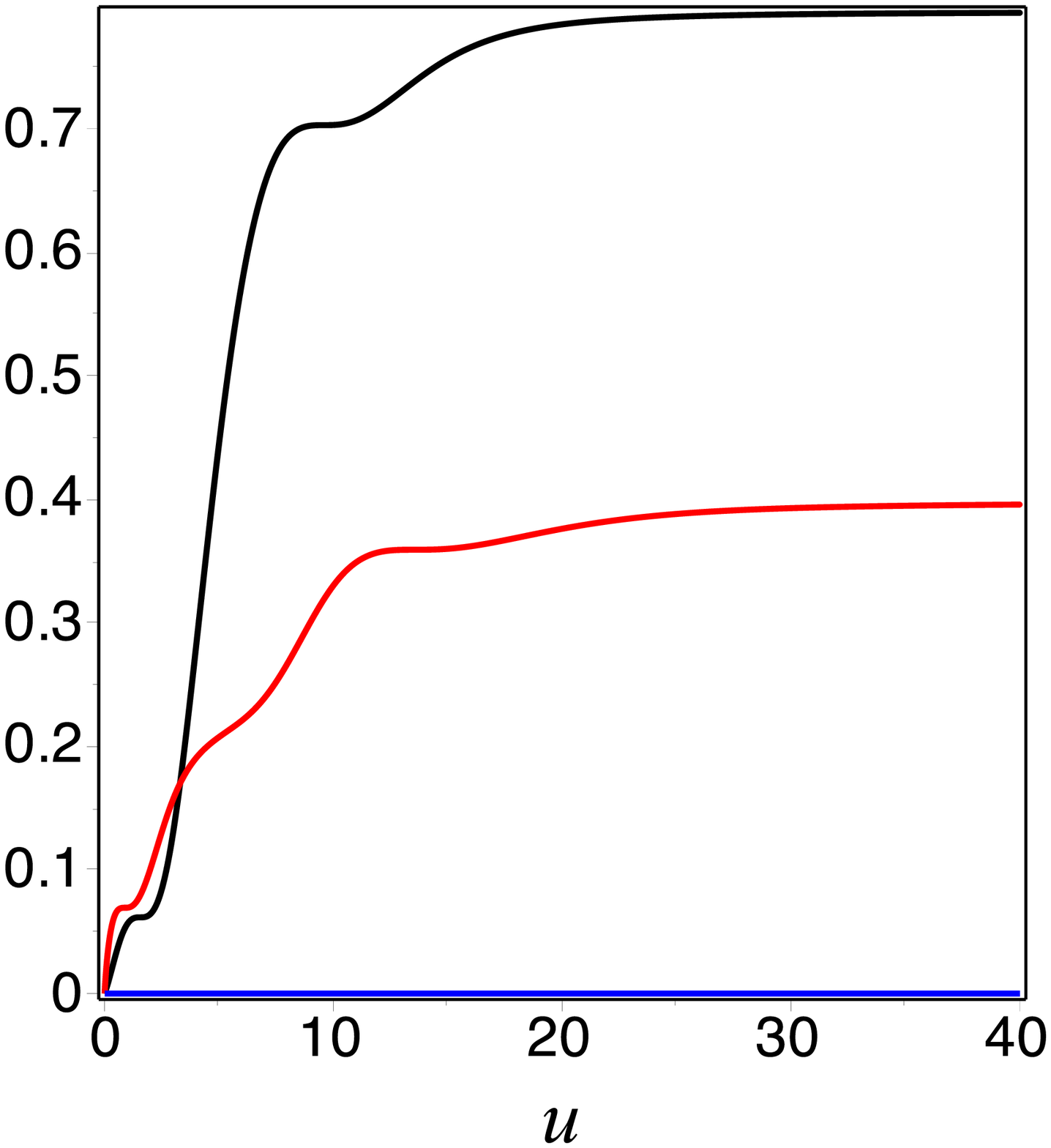}
\end{center}
{\renewcommand{\baselinestretch}{1}
\caption{Behavior of $I_\psi(u)$ (black line), $I_2(u)$ (red line) and $I_3(u)$ (blue line) in which $\psi_0(\rho),A_{03}(\rho)=0$. Here $A_{02}(y;p)=p \mathrm{e}^{-(y-1)^2}$, where for the sake of convenience we have introduced the transformation $\rho=y^2$. The plots from the left to right correspond to $p=0.9,1.06,1.10,1.13$. Notice the accentuated increase of mass extraction by the mode $\psi$ due to the enhancement effect when $p>p_*$ with $p_* \approx 1.06$.}}
\end{figure}

At this point, we are in conditions of undercover an important aspect of the interaction of electromagnetic and gravitational waves by focusing in the extraction of mass through each channel provided by each wave mode $\psi$, $A_2$ and $A_3$. In this instance, we have evolved numerically the wave modes starting with the initial data $\psi(u_0,\rho)=\psi_0(\rho)$, $A_{2}(u_0,\rho)=A_{02}(\rho)$ and $A_{3}(u_0,\rho)=A_{03}(\rho)$ \cite{celestino}. The mass of the system, in turn, is determined by the initial data, and according to with the Bondi formula (4) decay due to the action of the gravitational and electromagnetic waves. We propose the first experiment in which $\psi_0(\rho)=A_{03}(\rho)=0$ and $A_{03}(\rho;p) \neq 0$, where $p$ is a parameter that controls the amplitude of the initial distribution of the potential. In this situation it can be shown that $A_3(u,\rho) = 0$ for all $u > u_0$, nevertheless due to the coupling of the potentials $\psi$ and $A_2$, it follows that $\psi(u,\rho) \neq 0$ for all $u > u_0$. It means that the electromagnetic wave produces polarized gravitational waves. We might quantify how the gravitational wave is excited by the amount of mass carried or, in other words, to look at the behavior of $I_\psi(u)$ together with $I_2(u)$ for several values of increasing $p$. In Fig. 1 we depict the sequence of plots of $I_\psi(u)$ and $I_2(u)$ for $p = $. We have noticed an interesting effect as $p$ is increased. For $p$ smaller than a particular value we call it critical value, $p=p_*$, $I_\psi(u)$ is always smaller than $I_2(u)$; for $p \approx p_*$ we have that $I_\psi(u) \approx I_2(u)$, but for $p > p_*$ it follows that the mass extracted by gravitational radiation experiences an acentuated increase which is translated by $I_\psi(u) > I_2(u)$ for most of $u > u_0$. We have named this feature as the \textit{enhancement effect}.  

\begin{figure}[h]
\begin{center}
\includegraphics[scale=0.2]{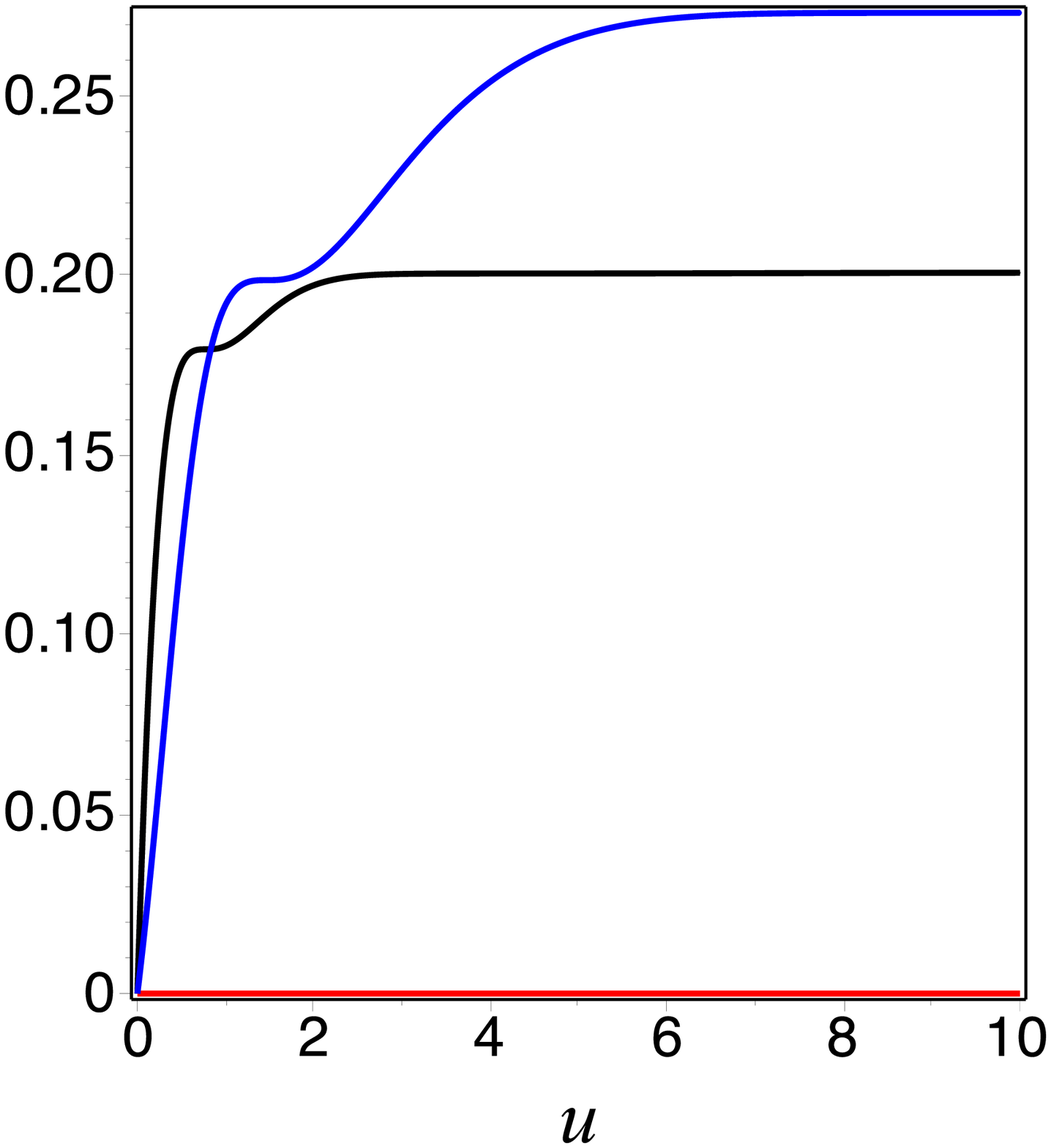}\includegraphics[scale=0.2]{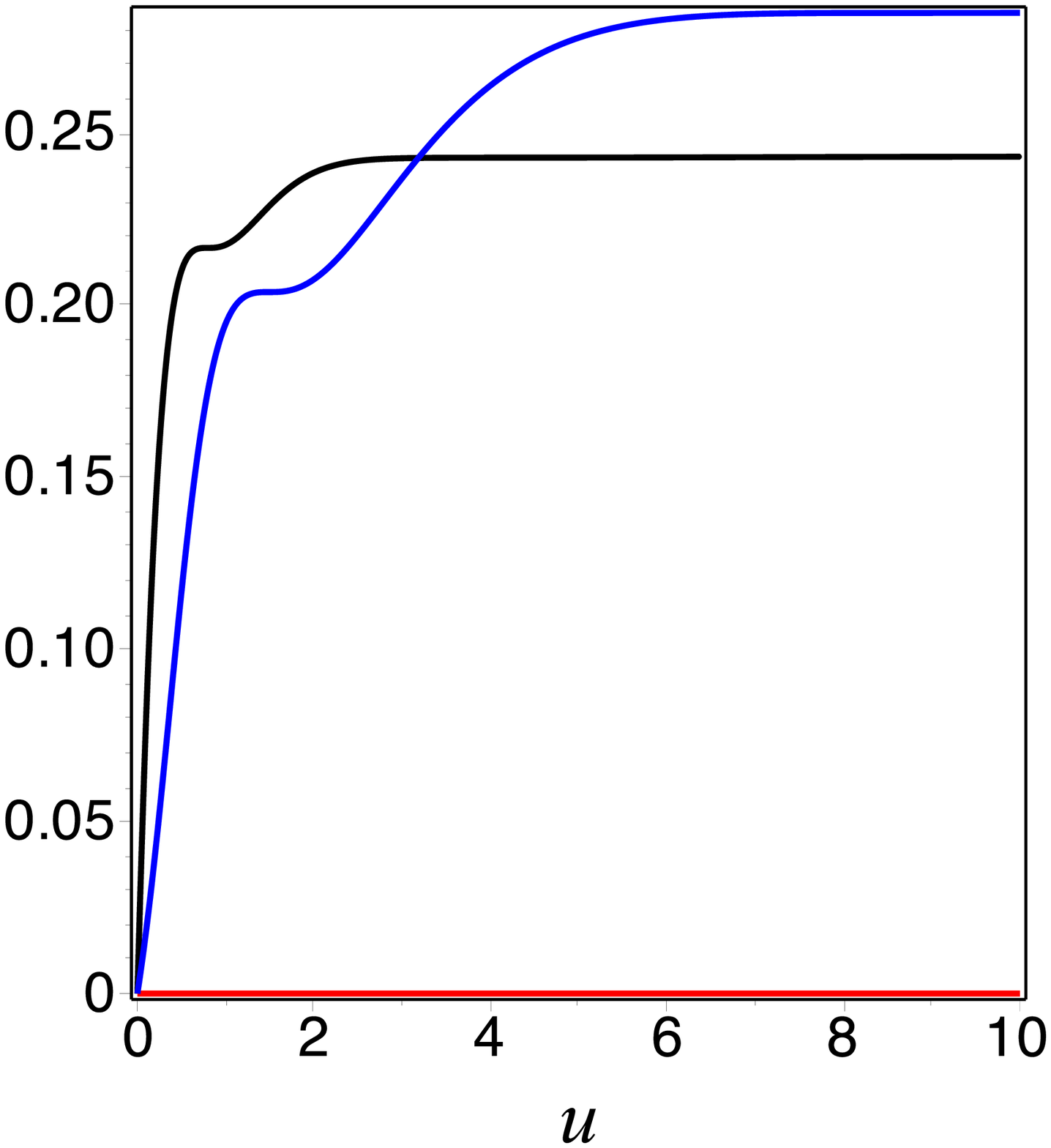}
\includegraphics[scale=0.2]{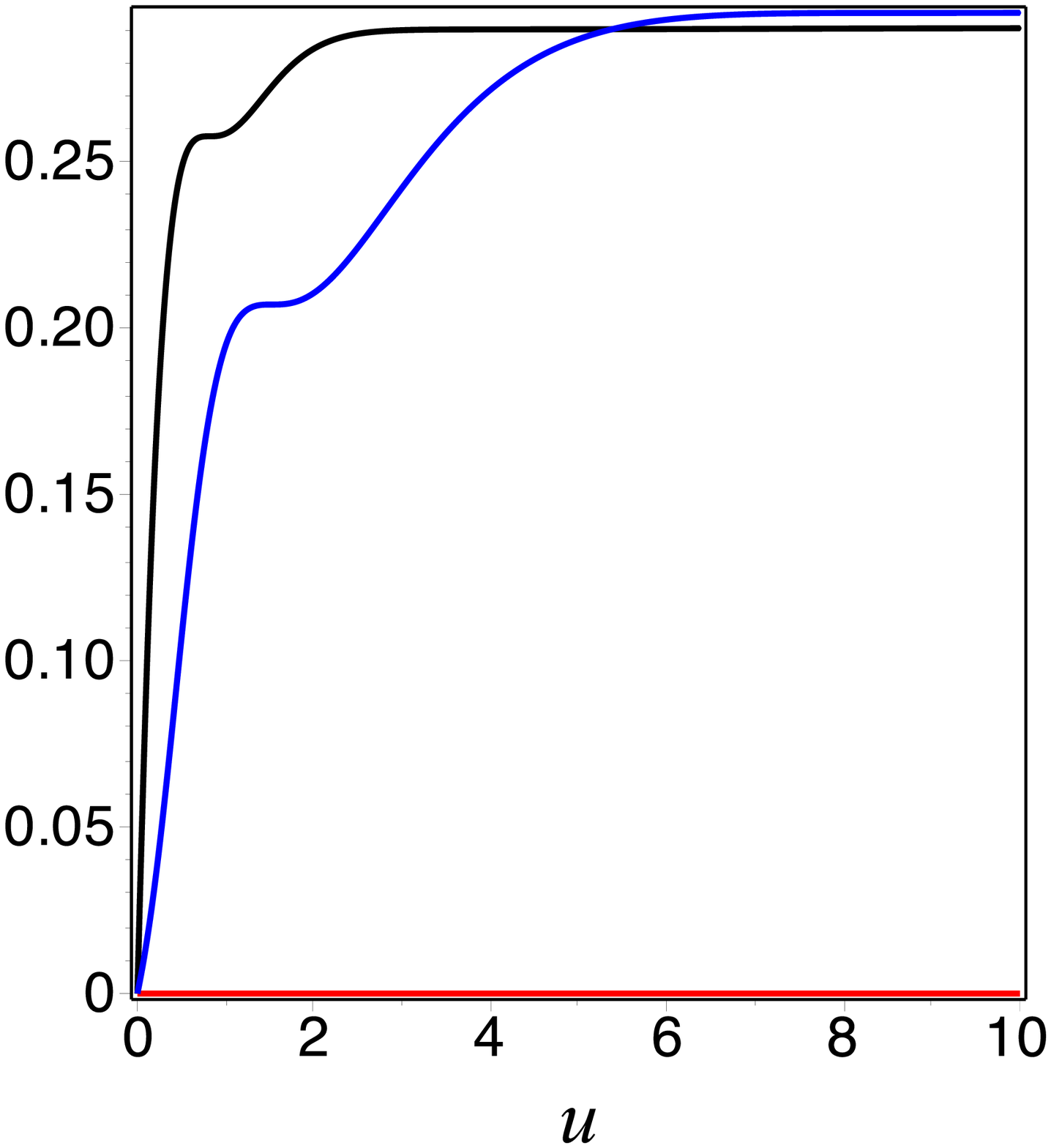}\includegraphics[scale=0.2]{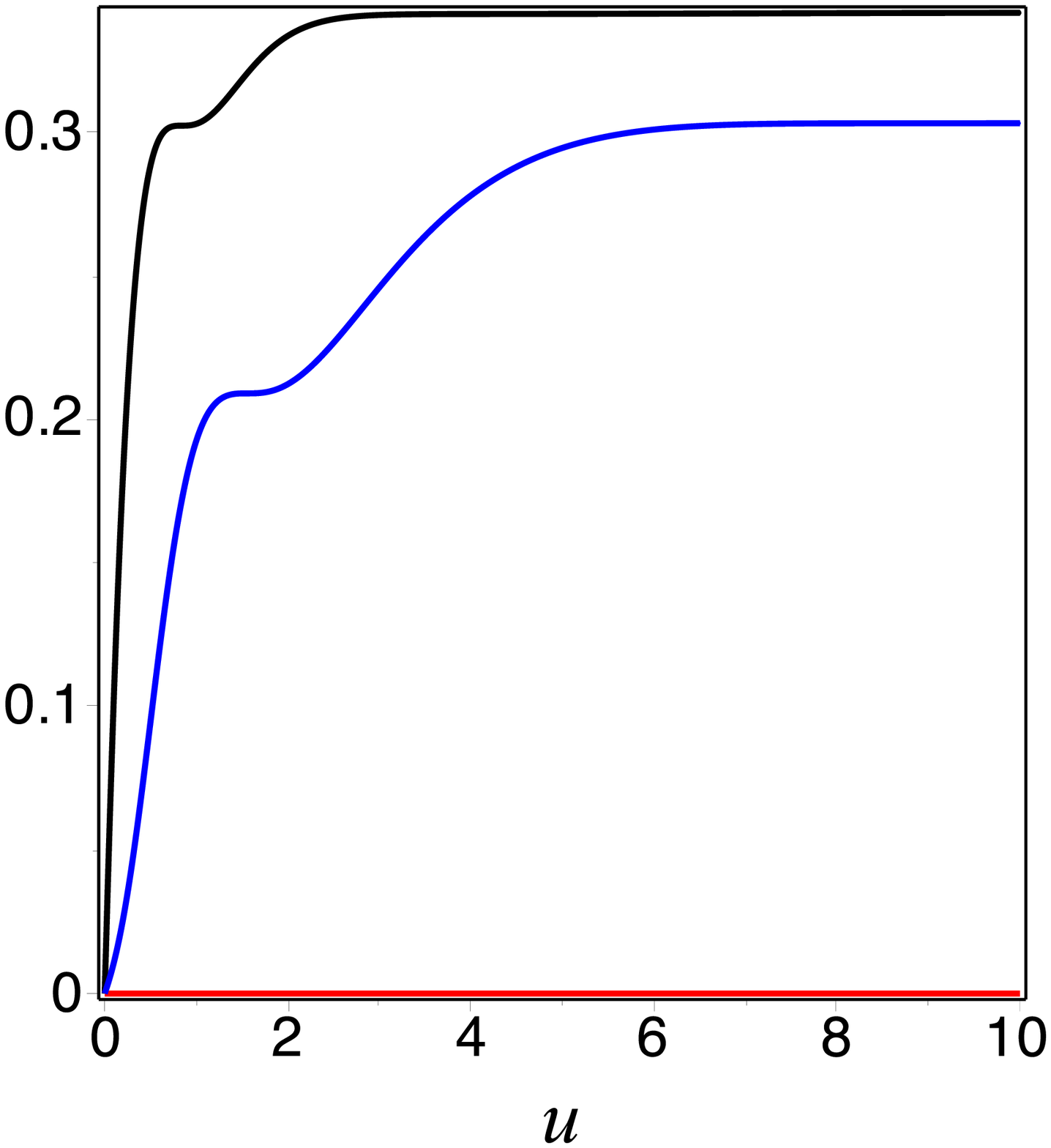}
\end{center}
{\renewcommand{\baselinestretch}{1}
\vspace{-0.2cm}
\caption{Behavior of $I_\psi(u)$ (black line), $I_2(u)$ (red line) and $I_3(u)$ (blue line) in which $\psi_0(\rho),A_{02}(\rho)=0$. Here $A_{03}(y;p)=p y^4\mathrm{e}^{-0.25(y-1/3)^2}/(1+y^2)$, where again $\rho=y^2$. The plots from the left to right correspond to $p=1.8,1.9,2.0,2.1$. Again, the enhancement effect takes place and $p_* \approx 2.0$.}}
\end{figure}

In the realm of interacting cylindrical gravitational and electromagnetic waves the enhancement seems to be generic. We have performed a further experiment with $\psi_0(\rho)=A_{02}(\rho)=0$ and $A_{03}(\rho;p) \neq 0$, and again, according with Fig. 2 the same behavior takes place. Finally, for the case where both electromagnetic potentials are taking part of the dynamics, i.e. for choosing initially $A_{02}(\rho),A_{03}(\rho) \neq 0$, we have noticed that there is a critical value $p_*$ such that $I_\psi(u)$ becomes of the same order of $I_2(u)$ and $I_3(u)$, and if $p > p_*$ the flow of mass through the gravitational channel is more accentuated than the mass extracted by the waves modes $A_2$ and $A_3$ (cf. Fig. 3). 

\begin{figure}[h]
\begin{center}
\includegraphics[scale=0.2]{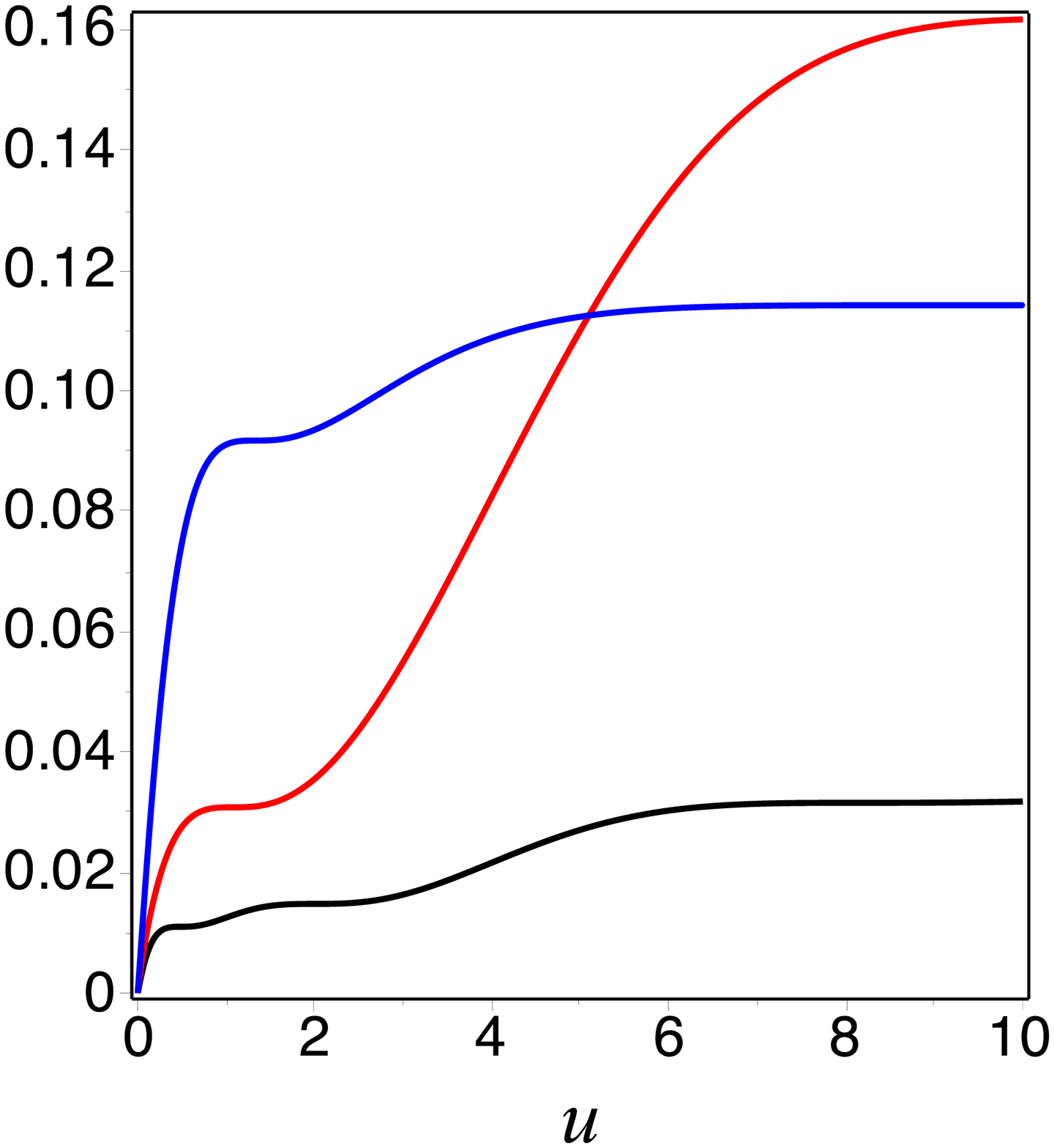}\includegraphics[scale=0.2]{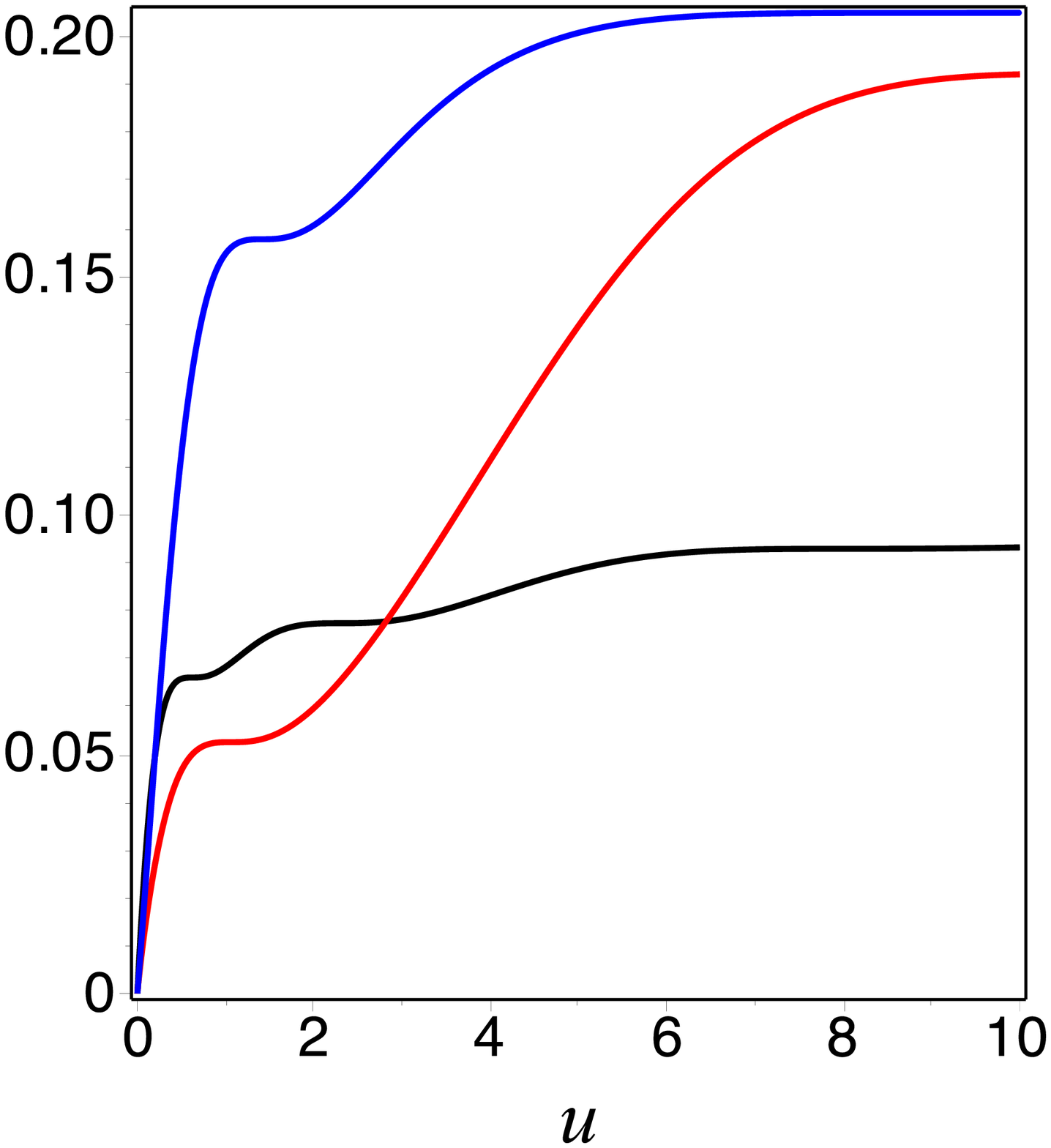}
\includegraphics[scale=0.2]{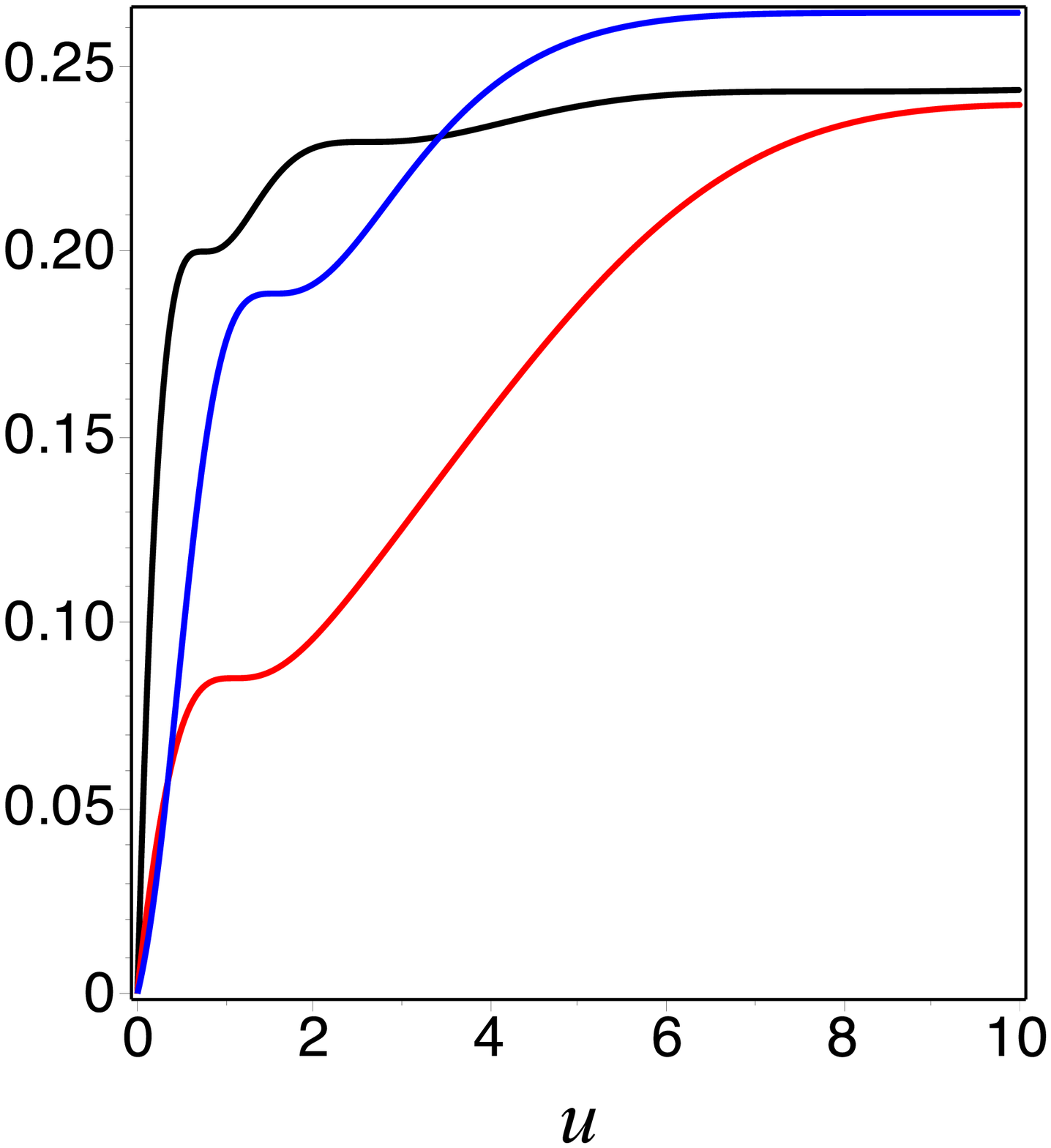}\includegraphics[scale=0.2]{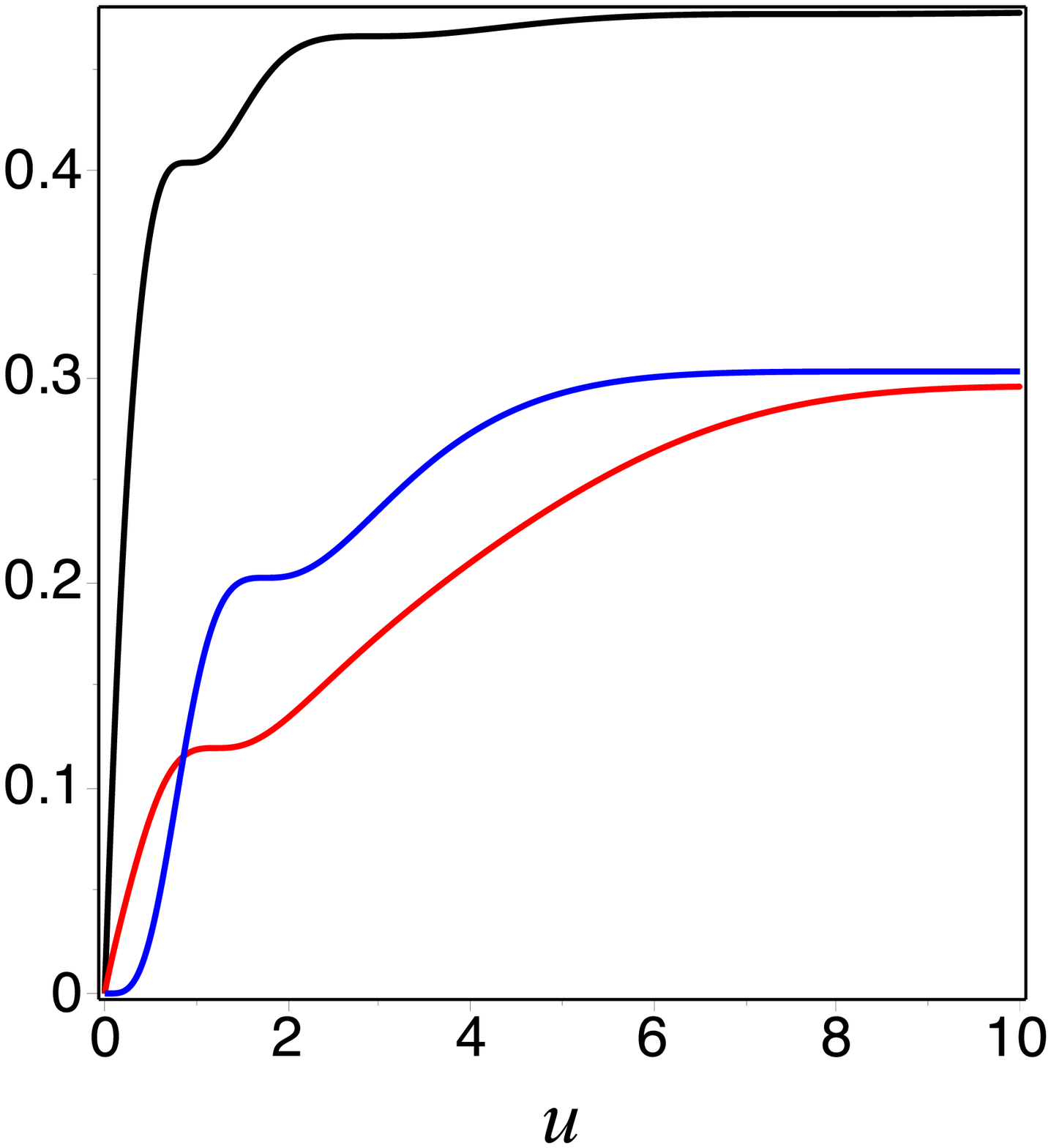}
\end{center}
{\renewcommand{\baselinestretch}{1}
\caption{Behavior of $I_\psi(u)$ (black line), $I_2(u)$ (red line) and $I_3(u)$ (blue line) in which only $\psi_0(\rho)=0$. We have the initial data $A_{02}(y;p)=0.5 \mathrm{e}^{-(y-1)^2}$ and $A_{03}(y;p)=p y^4\mathrm{e}^{-0.25(y-1/3)^2}/(1+y^2)$, where $\rho=y^2$. The plots from the left to right correspond to $p=1.0,1.5,2.0,2.5$. The enhancement of the gravitational mode occurs but the value of $p_*$ depends on the combination of the initial amplitudes of the electromagnetic potentials.}}
\end{figure}

\begin{acknowledgements}
The authors thank the financial support of Brazilian agencies CNPq and FAPERJ. HPO thanks FAPERJ for support within the grant BBP (Bolsas de Bancada para Projetos). 
\end{acknowledgements}

\end{document}